\documentclass{article}
\usepackage{amssymb}
\usepackage{amsfonts}
\usepackage{amsmath}

\input{tcilatex}
\begin{document}

\title{The hybrid meson: new results from the updated $m_{g}$ and $\alpha
_{s}$ parameters.}
\author{F. Iddir\thanks{%
E-mail: $iddir@univ$-$oran.dz$} \ and L. Semlala\thanks{%
E-mail: $l\_semlala@yahoo.fr$} \\
Laboratoire de Physique Th\'{e}orique\\
Universit\'{e} d'Oran Es-S\'{e}nia, ALGERIA}
\maketitle

\begin{abstract}
We present new results concerning the masses and the decay widths of the
most interesting hybrid meson states taking as inputs the gluon mass $m_{g}$
and the non-perturbative QCD running coupling constant $\alpha _{s}(0)$
comming from both LQCD and SDE recent estimations.
\end{abstract}

\section{Introduction}

Hybrid meson is one of the most promising new species of hadrons allowed by
QCD and subject of lot of works both in the theoretical and experimental
levels.

The hybrid mesons are studied from different models: lattice QCD$^{\text{%
\cite{LQCD_1},\cite{LQCD_2}}}$, flux tube model$^{\text{\cite{FluxTube_1}-%
\cite{FluxTube_6}}}$, bag model$^{\text{\cite{Bag_1}-\cite{Bag_7}}}$, QCD
sum rules$^{\text{\cite{QCD-SR_1}-\cite{QCD-SR_13}}}$, constituent gluon
models$^{\text{\cite{Orsay_1}-\cite{Kalash1} }}$ and from Effective
Hamiltonian model$^{\text{\cite{800 QCD-Hamilt_2}-\cite{600 QCD-Hamilt_2}}}$%
. Some of them can perform both estimations of masses and decay widths.

The nature of gluonic field inside hybrid is not yet be clear because the
gluon plays a double role: it propagates the interaction between color
sources and being itself colored it undergoes the interaction. Whereas, LQCD
and QCD Sum rules make no assumptions about it, two important hypothesis can
be retained from literature. The first one consider gluonic degrees of
freedom as \textquotedblright excitations\textquotedblright\ of the
\textquotedblright flux tube\textquotedblright\ between quark and antiquark,
which leads to the linear potential, that is familiar from quark model
(flux-tube model).

The second issue, which are supported by the present work, assumes that
hybrid is a bound state of quark-antiquark and a constituent glue which
interact through an adequate phenomenological potential. We can adapt our
interaction scheme with the idea of \textit{confined} and \textit{confining}
gluons$^{\text{\cite{Buttner}}}$( In the Landau and Coulomb gauges and in
interpolating gauges between them).

A confining gluons establish an area law behavior of the Wilson loop and the
linearly rising interquark confinement. the \textit{confined} gluons do not
propagate over long distances. We can accommodate \textit{confined}
(massive, constituent) gluon in coexistence with an effective quark
interaction which is confining$^{\text{\cite{Alkofer}}}$.

The quark model is still necessary for describing most of the available
spectroscopic data and their free parameters are fitted to the experiment.
In addition, our generalized Quark Model with Constituent Glue$^{\text{\cite%
{wari:IJMPA2008}}}$ uses two free parameters: the constituent gluon mass $%
m_{g}$\ and the decay parameter $\alpha _{s}$(the effective quark-gluon
vertex coupling). They must be set otherwise, since no (confirmed) hybrid
mesons data are available.

Several studies$^{\text{\cite{Alkofer}, \cite{Cornwall1}-\cite{Cornwall3}}}$
support the hypothesis that the gluon may develop a dynamical mass which is
intrinsically related to an infrared finite gluon propagator$^{\text{\cite%
{Aguilar1}}}$. This is consistent with QCD lattice simulations$^{\text{\cite%
{Bonnet}-\cite{LQCD:Oliveira}}}$. From the theoretical point of view, a
non-vanishing gluon mass is welcome to regularize infrared divergences and
solve some problems related with unitarity. From the phenomenological point
of view, a non-vanishing gluon mass is welcome by diffractive phenomena$^{%
\text{\cite{Forshaw}}}$ and inclusive radiative decays of J/$\Psi $and $%
\gamma ^{\text{\cite{Field}}}$. For, the glueball states, colorsinglet bound
states of gluons, are considered to be fairly massive e.g., about 1.5 $GeV$
for the lowest 0$^{++}$ and about 2 GeV for the lowest 2$^{++}$, as
indicated in lattice QCD calculations$^{\text{\cite{Rothe},\cite{Ishii}}}$,
a simple constituent gluon picture\ may be approximately obtained as $%
M_{GB}\simeq 2m_{g}$ for the glueball mass $M_{GB}$.

There are many theoretical evidences that the QCD effective charge $\alpha
_{s}$ freezes at small momenta$^{\text{\cite{Cornwall1},\cite{Cornwall2}, 
\cite{Binosi}-\cite{Natale}}}$. The infrared finiteness of the effective
charge can be considered as one of the manifestation of the phenomenon of
dynamical gluon mass generation$^{\text{\cite{Cornwall1},\cite{Cornwall2},%
\cite{Aguilar1}}}$

Phenomenology sensitive to infrared properties of QCD gives$^{\text{\cite%
{Luna}-\cite{Aguilar4}}}$ $\alpha _{s}(0)$ $\simeq $ $0.7\pm \ 0.3$. The
phenomenological evidences for the strong coupling constant freezing in the
infrared (IR) are much more numerous. Models where a static potential is
used to compute the hadronic spectra make use of a frozen coupling constant
at long distances$^{\text{\cite{Eichten}-\cite{Godfrey}}}$.

\section{Update the gluon mass}

In earlier estimations$^{\text{\cite{wari:IJMPA2008}}}$ we have used $%
m_{g}\sim 800$ $MeV$ to the best fit of our results to the ones obtained by
lattice calculations of $1^{-+}c\overline{c}g$ and $b\overline{b}g$ masses.
This is also compatible with other works (table 1).%
\begin{equation*}
\underset{\text{Table 1. Some estimations of the }m_{g}\text{ around }800%
\text{ }MeV}{%
\begin{tabular}{|c|c|c|}
\hline
Authors & Method & $m_{g}$ \\ \hline
Parisi, Petronzio$\text{\cite{Parisi}}$ & J/$\psi \rightarrow \gamma $X & 
800 MeV \\ \hline
Donnachie, Landshoff$\text{\cite{Donnachie}}$ & Pomeron parameters & 687-985
MeV \\ \hline
Hancock, Ross$\text{\cite{Hancock}}$ & Pomeron slope & 800 MeV \\ \hline
Nikolaev et al.$\text{\cite{Nikolaev}}$ & Pomeron parameters & 750 MeV \\ 
\hline
Spiridonov, Chetyrkin$\text{\cite{Spiridonov}}$ & $\Pi _{\mu \nu }^{em},%
\text{ }\langle TrG_{\mu \nu }^{2}\rangle $ & 750 MeV \\ \hline
Field$\text{\cite{Field2}}$ & $J/\psi \rightarrow \gamma X$ & 0.721$%
_{-0.068}^{+0.016}$GeV \\ \hline
\end{tabular}%
}
\end{equation*}

Recent lattice data$^{\text{\cite{LQCD:Takumi},\cite{LQCD:Oliveira}}}$
estimate $m_{g}$ around $600$ $MeV$. This value has been already obtained
many years ago in the context of Schwinger-Dyson equations (SDEs)$^{\text{%
\cite{Cornwall1}}}$, this is also compatible with the effective QCD Coulomb
gauge Hamiltonian approach$^{\text{\cite{600 QCD-Hamilt_1}},\text{\cite{600
QCD-Hamilt_2}}}$. We choose this value as\ an input for our model, described
by the binding Hamiltonian%
\begin{equation}
H=\sum\limits_{i=q,\text{ }\bar{q},\text{ }g}\left( \frac{\vec{p}_{i}^{\text{
}2}}{2M_{i}}+\frac{M_{i}}{2}+\frac{m_{i}^{2}}{2M_{i}}\right) +V_{eff\text{ }}
\end{equation}

$V_{eff}$ is the average over the color space of chromo-spatial potential: 
\begin{eqnarray}
V_{eff} &=&\left\langle V\right\rangle _{color}=\left\langle
-\sum\limits_{i<j=1}^{N}\mathbf{F}_{i}\cdot \mathbf{F}_{j}\text{ }%
v(r_{ij})\right\rangle _{color}  \notag \\
&=&\sum\limits_{i<j=1}^{N}\alpha _{ij}v(r_{ij})\text{ };
\end{eqnarray}%
where $v(r_{ij})$ is the phenomenological potential term.

We take a potential which has the form: 
\begin{equation}
v(r_{ij})=v_{CL}(r_{ij})+v_{SD}(r_{ij});
\end{equation}%
where the QCD-motivated \textquotedblright
Coulomb.+Linear\textquotedblright\ term reads: 
\begin{equation}
v_{CL}(r_{ij})=-\frac{\alpha _{s}}{r_{ij}}+\sigma \text{ }r_{ij}+c\text{ }.
\end{equation}

The spin-dependent term can split into Spin-Spin, Spin-Orbit and Tensor
terms : 
\begin{equation}
v_{SD}=v_{SS}+v_{SO}+v_{T};
\end{equation}

\begin{equation}
\left( v_{SS}\right) _{ij}=\QDATOP{\frac{8\pi \alpha _{h}}{3M_{i}M_{j}}\frac{%
\sigma _{h}^{3}}{\sqrt{\pi ^{3}}}\exp (-\sigma _{h}^{2}\text{ }r_{ij}^{2})%
\text{ }}{\frac{8\pi \alpha _{s}}{3M_{i}M_{j}}\delta ^{3}\left( \mathbf{r}%
_{ij}\right) \text{ \ \ \ \ \ \ \ \ \ \ \ \ \ }}\QDATOP{\mathbf{s}_{i}\cdot 
\mathbf{s}_{j}}{\mathbf{s}_{i}\cdot \mathbf{s}_{j}}\QDATOP{\text{(light
sector) \ \ }}{\text{(charm sector) \ }};
\end{equation}

\begin{eqnarray}
\left( v_{SO}\right) _{ij} &=&\frac{\alpha _{S}}{2r_{ij}^{3}}(\frac{\mathbf{s%
}_{i}\cdot \mathbf{r}_{ij}\times \mathbf{p}_{i}}{M_{i}^{2}}-\frac{\mathbf{s}%
_{j}\cdot \mathbf{r}_{ij}\times \mathbf{p}_{j}}{M_{j}^{2}}  \notag \\
&&-\frac{2\mathbf{s}_{i}\cdot \mathbf{r}_{ij}\times \mathbf{p}_{j}-2\mathbf{s%
}_{j}\cdot \mathbf{r}_{ij}\times \mathbf{p}_{i}}{M_{i}M_{j}});
\end{eqnarray}%
\begin{equation}
\left( v_{T}\right) _{ij}=\frac{\alpha _{S}}{M_{i}M_{j}r_{ij}^{3}}(3\text{ }%
\mathbf{s}_{i}\cdot \mathbf{\hat{r}}_{ij}\text{ }\mathbf{s}_{j}\cdot \mathbf{%
\hat{r}}_{ij}-\mathbf{s}_{i}\cdot \mathbf{s}_{j}).
\end{equation}

Using the variational method, one can find the mass and the wavefunction of
any $J^{PC}$ hybrid state$^{\text{\cite{wari:IJMPA2008}}}$.

The energy of the constituent gluon inside the hybrid $\omega \simeq M_{g}$
can be evaluated using the condition%
\begin{equation}
\frac{\partial E}{\partial M_{i}}=0\text{ }.
\end{equation}

The value of $\omega $\ and the hybrid wavefunctions are used to evaluate
the spatial overlaps in the decay width calculations.

\section{Update the decay width parameters}

The decay of an hybrid state A into two ordinary mesons B and C is
represented by the matrix element of the Hamiltonian annihilating a gluon
and creating a quark pair (QPC model)$^{\text{\cite{Orsay_1}}}$: 
\begin{equation}
\langle BC\left\vert H\right\vert A\rangle =gf(A,B,C)\left( 2\pi \right)
^{3}\delta _{3}\left( p_{A}-p_{B}-p_{C}\right) ;
\end{equation}%
where $f(A,B,C)$ is the decay amplitude involving the flavor, the color, the
spatial and the (non-relativistic) spin overlaps. The spatial overlap is
given by: 
\begin{eqnarray}
I &=&\iint \frac{d\overrightarrow{p}d\overrightarrow{k}}{\left( 2\pi \right)
^{6}\sqrt{2\omega }}\Psi _{q\overline{q}g}^{l_{q\overline{q}}m_{q\overline{q}%
}l_{g}m_{g}}\left( \overrightarrow{P}_{B}-\overrightarrow{p},\overrightarrow{%
k}\right)  \notag \\
&&\times \Psi _{q_{i}\overline{q}}^{l_{B}m_{B}\ \ast }\left( \overrightarrow{%
p}_{1}\right) \Psi _{q\overline{q}_{i}}^{l_{C}m_{C}\ \ast }\left( 
\overrightarrow{p}_{2}\right) Y_{l}^{m\ \ast }\left( \Omega _{B}\right)
d\Omega _{B},  \label{overlap}
\end{eqnarray}

and the partial width by: 
\begin{equation}
\Gamma \left( A\rightarrow BC\right) =4\alpha _{s}\left\vert f\left(
A,B,C\right) \right\vert ^{2}\frac{P_{B}E_{B}E_{C}}{M_{A}};
\end{equation}

where $\alpha _{s}$ represents the infrared quark-gluon vertex coupling. We
have always chosen $\alpha _{s}\approx 1$ (and $\omega \approx 0.8$ $GeV$)
in our previous works, but now we can use a more convincing values available
from recent Schwinger-Dyson equations (SDEs) and Lattice QCD data.

There are two characteristic definitions of the effective charge, frequently
employed in the literature. The first definition is obtained within the
pinch technique (PT) framework$^{\text{\cite{Cornwall1}-\cite{Cornwall2}}}$%
(which can be appropriately extended to the Taylor ghost-gluon coupling$^{%
\text{\cite{Aguilar5},\cite{pepe}}}$).This effective charge, to be denoted
by $\alpha _{PT}$, constitutes the most direct non-abelian generalization of
the familiar concept of the QED effective charge. The second definition of
the QCD effective charge, to be denoted by $\alpha _{gh}$, involves the
ghost and gluon self-energies, in the Landau gauge, and in the kinematic
configuration where the well-known Taylor non-renormalization theorem
becomes applicable. $\alpha _{gh}$ has been employed extensively in lattice
studies (see for instance \cite{Bloch}-\cite{Boucaud2} and references
therein), where the Landau gauge is the standard choice for the simulation
of the gluon and ghost propagators, as well as in various investigations
based on Schwinger-Dyson equations (SDEs)$^{\text{\cite{Smekal}-\cite%
{Boucaud3}}}$. The two charges are identical not only in the deep UV, where
asymptotic freedom manifests itself, but also in the deep IR, where they
\textquotedblleft freeze\textquotedblright\ at the same non-vanishing value$%
^{\text{\cite{Aguilar5}}}$.

Using recent (quenched) lattice data on the gluon and ghost propagators, as
well as the Kugo-Ojima function, authors of reference \cite{Aguilar6}
extract the non-perturbative behavior of QCD effective charges.

They have offered a plausible explanation for the observed discrepancy in
the freezing values of the effective charges obtained from the lattice ($%
\alpha _{s}(0)\sim 2-2.5^{\text{\cite{alpha lattice1},\cite{alpha lattice2}}%
} $)\ and those derived from the fitting of various QCD processes, sensitive
to non-perturbative physics ($\sim 0.7\pm 0.3$). They claim that the
underlying reason for the discrepancy is the difference in the gauges
(Landau vs Feynman) used in the two approaches.

Since our decay model is obtained in the Feynman gauge$^{\text{\cite{Orsay_1}%
}}$, it's natural to choose $\alpha _{s}\simeq \alpha _{PT}(0)$
corresponding to the pinch technique gluon propagator, i.e. the background
field propagator calculated in the Feynman gauge. For $m_{g}=600$ $MeV$ We
have $\alpha _{s}\simeq \alpha _{PT}(0)\simeq $ $0.85^{^{\text{\cite%
{Aguilar6}}}}$.

We present in table 2, the\ updated hybrid masses using the recent LQCD
constituent gluon mass $m_{g}=600$ $MeV$. The gluon energy $\omega $
appearing in the decay's formulas, can be set to $M_{g}\simeq 1$ $GeV$, the
energy of the confined gluon.%
\begin{equation*}
\underset{\text{Table2. 1}^{-+}\text{ light and 1}^{--}\text{ charmed hybrid
mesons masses.}}{%
\begin{tabular}{|l|l|l|l|}
\hline
& Mass & $M_{q}$ & $M_{g}\simeq \omega $ \\ \hline
1$^{-+}n\bar{n}g$ & 1.70 & 0.87 & 1.08 \\ \hline
1$^{--}c\bar{c}g$ & 4.10 & 1.95 & 1.00 \\ \hline
\end{tabular}%
}
\end{equation*}

Since $\Gamma \left( A\rightarrow BC\right) \sim \frac{\alpha _{s}}{\omega }$%
, the old decay widths must be multiplied by the factor:%
\begin{equation*}
\frac{\left( \alpha _{s}\right) _{\text{New}}}{\left( \alpha _{s}\right) _{%
\text{Old}}}\frac{\left( \omega \right) _{\text{Old}}}{\left( \omega \right)
_{\text{New}}}=0.85\cdot 0.8=0.68
\end{equation*}

The results are summarized in tables 3-8.

\begin{equation*}
\underset{\text{.}}{\underset{\text{Table 3. \ \ Partial decay widths of the
(}M=1.6\text{) hybrid in (S+S)-standard mesons}}{%
\begin{tabular}{|l|l|}
\hline
$\Gamma _{\rho \pi }$ & $%
\begin{tabular}{ll}
36 & $MeV$%
\end{tabular}%
$ \\ \hline
$\Gamma _{K^{\ast }K}$ & $%
\begin{tabular}{ll}
10 & $MeV$%
\end{tabular}%
$ \\ \hline
$\Gamma _{\rho \omega }$ & $%
\begin{tabular}{ll}
1 & $MeV$%
\end{tabular}%
$ \\ \hline
\end{tabular}%
}}
\end{equation*}%
\begin{equation*}
\underset{\text{Table 4. \ Partial decay widths of the (}M=2.0\text{) hybrid
in (S+S)-standard mesons.}}{%
\begin{tabular}{|l|l|}
\hline
$\Gamma _{\rho \pi }$ & $%
\begin{tabular}{ll}
47 & $MeV$%
\end{tabular}%
$ \\ \hline
$\Gamma _{K^{\ast }K}$ & $%
\begin{tabular}{ll}
36 & $MeV$%
\end{tabular}%
$ \\ \hline
$\Gamma _{\rho \omega }$ & $%
\begin{tabular}{ll}
21 & $MeV$%
\end{tabular}%
$ \\ \hline
$\Gamma _{\rho (1450)\pi }$ & $%
\begin{tabular}{ll}
12 & $MeV$%
\end{tabular}%
$ \\ \hline
$\Gamma _{K^{\ast }(1410)K}$ & $%
\begin{tabular}{ll}
3 & $MeV$%
\end{tabular}%
$ \\ \hline
\end{tabular}%
}
\end{equation*}%
\begin{equation*}
\underset{\text{Table 5. \ Partial decay widths of the (}M=1.6\text{) hybrid
in (L+S)-standard mesons (in }MeV\text{).}}{%
\begin{tabular}{|c|c|c|c|}
\hline
$L$ & $0$ & $1$ & $2$ \\ \hline
$\Gamma _{b_{1}^{0}\pi ^{-}}\approx \Gamma _{b_{1}^{-}\pi ^{0}}$ & 98 & 294
& 491 \\ \hline
$\Gamma _{f_{1}^{0}(1285)\pi ^{-}}$ & 79 & 60 & 100 \\ \hline
$\Gamma _{f_{1}^{0}(1420)\pi ^{-}}$ & 26 & 20 & 33 \\ \hline
\end{tabular}%
\ }
\end{equation*}%
\begin{equation*}
\underset{\text{Table 6. \ Partial decay widths of the (}M=2.0\text{) hybrid
in (P+S)-standard mesons (in }MeV\text{)}}{%
\begin{tabular}{|c|c|c|c|}
\hline
$L$ & $0$ & $1$ & $2$ \\ \hline
$\Gamma _{b_{1}^{0}\pi ^{-}}\approx \Gamma _{b_{1}^{-}\pi ^{0}}$ & 106 & 317
& 528 \\ \hline
$\Gamma _{f_{1}^{0}(1285)\pi ^{-}}$ & 96 & 73 & 122 \\ \hline
$\Gamma _{f_{1}^{0}(1420)\pi ^{-}}$ & 67 & 51 & 85 \\ \hline
\end{tabular}%
\ }
\end{equation*}%
\begin{equation*}
\underset{\text{Table 7: Partial decay widths of the (}M=4.26\text{) hybrid
in (S+S)-standard mesons (in }MeV\text{).}}{%
\begin{tabular}{|c|c|c|c|}
\hline
$L$ & $0$ & $1$ & $2$ \\ \hline
$\Gamma _{D^{0}\bar{D}^{0}}$ & 88 & 264 & 440 \\ \hline
$\Gamma _{D^{+}D^{-}}$ & 92 & 276 & 460 \\ \hline
$\Gamma _{D_{s}^{+}D_{s}^{-}}$ & 97 & 291 & 486 \\ \hline
$\Gamma _{D^{\ast +}D^{\ast -}=}\Gamma _{D^{\ast 0}\overline{D^{\ast 0}}}%
\text{\ }%
\begin{tabular}{l}
$S=0$ \\ 
$S=1$ \\ 
$S=2$%
\end{tabular}%
\ $ & 
\begin{tabular}{l}
21 \\ 
\multicolumn{1}{c}{0} \\ 
34%
\end{tabular}
& 
\begin{tabular}{l}
63 \\ 
\multicolumn{1}{c}{0} \\ 
251%
\end{tabular}
& 
\begin{tabular}{l}
1 \\ 
\multicolumn{1}{c}{0} \\ 
17%
\end{tabular}
\\ \hline
\end{tabular}%
\ }
\end{equation*}

\begin{equation*}
\underset{\text{Table 8.\ Partial decay widths of the (}M=4.3\text{) hybrid
in (P+S)-standard mesons (in }MeV\text{).}}{%
\begin{tabular}{|c|c|}
\hline
$\Gamma _{D_{1}(2420)\overline{D^{0}}}=\Gamma _{\overline{D_{1}}(2420)D^{0}}$
$\simeq \Gamma _{D_{1}^{\pm }(2420)D^{\mp }}=\Gamma _{D_{1}^{\mp
}(2420)D^{\pm }}\sim $ & 78$=$312$\cdot \frac{1}{4}$ \\ \hline
\end{tabular}%
\ }
\end{equation*}

Note that LQCD$^{^{\text{\cite{LQCD 2200 decay}}}}$ gives the $2.2(2)$ $GeV$
hybrid partial decay widths: $\Gamma _{b_{1}\pi }=400\pm 120$ $MeV$\ $>$ $%
\Gamma _{f_{1}\pi }=90\pm 60$ $MeV$ which are in agreement with our results.
The experimental results for $\pi _{1}(2000)$ are: $\Gamma _{b_{1}\pi
}=230\pm 32\pm 73$ $MeV^{\text{\cite{Data1 pi(2000)}}}\lesssim $ $\Gamma
_{f_{1}\pi }=333\pm 52\pm 49$ $MeV^{\text{\cite{Data2 pi(2000)}}}$, It's
difficult to reconciliate our partial decay width $\Gamma _{1^{-+}n\overline{%
n}g(2000)->f_{1}\pi }$ with the experimental one.

The same remark holds for $\Gamma _{1^{-+}n\overline{n}g(1600)->f_{1}\pi }$
which disagrees with the experimental data $240\pm 60$ $MeV^{\text{\cite%
{Data1 pi(1600)}}}$ and for $\Gamma _{1^{-+}n\overline{n}g(1600)->\rho \pi }$%
which is very far from the experimental values $269\pm 21$ $MeV^{\text{\cite%
{Data2 pi(1600)}}}$ and $168\pm 20$ $MeV^{\text{\cite{Data3 pi(1600)}}}$.

\section{Conclusion}

In the framework of the quark model with constituent glue, and taking into
account the new values of $m_{g}$ and $\alpha _{s}(0)$ parameters, available
from recent LQCD and DSE calculations, we have updated our previous
estimations$^{\text{\cite{wari:IJMPA2008}}}$ of masses and decay widths of
the more interesting hybrid meson states.

We found that $M_{1^{-+}n\bar{n}g}\sim 1.7$ $GeV$ and the decay widths $%
\Gamma _{1^{-+}n\overline{n}g(1600)->\rho \pi }$ and $\Gamma _{1^{-+}n%
\overline{n}g(1600)->f_{1}\pi }$ are in disagreement with the experimental
ones. So we disregard the hypothesis of the hybrid meson structure of the
candidate $\pi _{1}(1600).$

The estimated mass of the $1^{-+}n\overline{n}g$\ GE-hybrid is around $1.9$ $%
GeV$, but we can't interpret this state as the candidate $\pi _{1}(2000)$,
because It's difficult to reconcile our partial decay widths with the
experimental results.

In the charm sector, the $1^{--}c\overline{c}g$ is estimated to have a mass
around $4.1$ $GeV$ which is consistent with the candidate Y(4260), and the
partial decay widths show that the $1^{--}c\overline{c}g$ can generate
observable resonances in both channels\ "S+S" and "S+P". Note that the
mixing of $1^{--}c\overline{c}g$ and the corresponding $1^{--}c\overline{c}$
is excluded$^{\text{\cite{Safir_2}}}$.

\section*{Acknowledgments}

This work is supported by the Laboratoire de Physique Th\'{e}orique d'Oran
Es-S\'{e}nia.

\newpage

\end{document}